\documentclass[sigconf,screen]{acmart}

\renewcommand\footnotetextcopyrightpermission[1]{} 
\usepackage{algorithmic}
\usepackage{graphicx}
\usepackage{textcomp}
\usepackage{xcolor}
\usepackage{stfloats}
\usepackage{url}
\usepackage{enumitem}
\usepackage{booktabs} 
\usepackage{multirow} 
\usepackage{color, colortbl}
\definecolor{Gray}{gray}{0.9}

\graphicspath{ares2021-153/source/}

\usepackage{pifont}

\usepackage{tikz} 

\newcommand*\circled[2][]{\tikz[baseline=(char.base)]{
    \node[shape=circle,draw,inner sep=0pt,#1,every number/.try] (char) {#2};}}
\tikzstyle{every number}=[draw=white] 

\newcommand{\RansomwareName}{\textsc{RansomClave}} 

\usepackage[ruled,vlined,noend]{algorithm2e} 
\usepackage[ruled,vlined,noend]{algorithm2e}

\newcommand{\tinyskip}{\vspace{3pt}}
\newcommand{\mypar}[1]{\tinyskip\noindent\textbf{#1.}\xspace}
\newcommand{\myparr}[1]{\tinyskip\noindent\textbf{#1}\xspace}

\newenvironment{myitemize}{%
\begin{itemize}[leftmargin=1em, itemsep=.1em, parsep=.1em, topsep=.1em,
    partopsep=.1em]}
{\end{itemize}}

\begin{document}

\title{\RansomwareName{}: Ransomware Key Management using SGX}

\author{Alpesh Bhudia} 
\affiliation{
\institution{Information Security Group \\ Royal Holloway, University of London \\
alpesh.bhudia.2018@live.rhul.ac.uk}
\city{}
\country{}
}

\author{Daniel O'Keeffe}
\affiliation{\institution{Royal Holloway, University of London  \\
daniel.okeeffe@rhul.ac.uk}
\city{}
\country{}
}

\author{Daniele Sgandurra}
\affiliation{
\institution{Information Security Group \\ Royal Holloway, University of London \\
daniele.sgandurra@rhul.ac.uk}
\city{}
\country{}
}

\author{Darren Hurley-Smith}
\affiliation{
\institution{Information Security Group \\ Royal Holloway, University of London \\
darren.hurley-smith@rhul.ac.uk}
\city{}
\country{}
}

\begin{abstract}
Modern ransomware often generate and manage cryptographic keys on the victim's machine, giving defenders an opportunity to capture exposed keys and recover encrypted data without paying the ransom. However, recent work has raised the possibility of future \emph{enclave-enhanced} malware that could avoid such mitigations using emerging support for hardware-enforced secure enclaves in commodity CPUs. Nonetheless, the practicality of such enclave-enhanced malware and its potential impact on all phases of the ransomware lifecyle remain unclear. Given the demonstrated capacity of ransomware authors to innovate in order to better extort their victims (e.g. through the adoption of untraceable virtual currencies and anonymity networks), it is important to better understand the risks involved and identify potential mitigations.  

As a basis for comprehensive security and performance analysis of enclave-enhanced ransomware, we present \RansomwareName{}, a family of ransomware that securely manage their cryptographic keys using an enclave. We use \RansomwareName{} to explore the implications of enclave-enhanced ransomware for the key generation, encryption and key release phases of the ransomware lifecycle, and to identify potential limitations and mitigations.

We propose two plausible victim models and analyse, from an attacker's perspective, how \RansomwareName{} can protect cryptographic keys from each type of victim. We find that some existing mitigations are likely to be effective during the key generation and encryption phases, but that \RansomwareName{} enables new \emph{trustless} key release schemes that could potentially improve attacker's profitability and, by extension, make enclaves an attractive target for future attackers.

\end{abstract}

\maketitle

\section{Introduction}
Over the last decade, ransomware has become one of the more prominent cyber-security threats against individuals and organisations~\cite{keshavarzi2020i2ce3}. Ransomware is a malware that either renders a victim's computer resources (\emph{locker-ransomware}) or valuable data (\emph{crypto-ransomware}) unusable \cite{luo2007awareness}. Crypto-ransomware is currently the most common of the two, and is especially difficult to defend against since victims are required to obtain the corresponding decryption key to access their data (assuming no backup is available).

A typical ransomware attack lifecycle goes through four main phases: installation, unique public/private key generation, encryption (with symmetric keys), and extortion/private key release. For a successful operation, the private key created in the second phase needs to be securely stored and only released in the last phase once the victim has paid the ransom
\cite{liska2016ransomware,al2018ransomware}. Previous work has shown several weaknesses in existing ransomware cryptographic key management systems~\cite{berrueta2019survey,davies2020evaluation,bajpai2018key,halderman2009lest}. For example, many ransomware variants (e.g., DMA Locker, Locky, Cerber, WannaCry, NotPetya, and BadRabbit) generate and store cryptographic keys (both asymmetric and symmetric ones) in untrusted memory on the victim's machine before permanently deleting them. Fortunately, victims can exploit such weaknesses to extract the generated keys using memory forensics techniques \cite{maartmann2009persistence} and tools like Dumpit, RAMCapturer and FTK imager~\cite{villalba2018ransomware}. Additionally, researchers have developed techniques, e.g. PayBreak \cite{kolodenker2017paybreak}, to capture keys generated at run-time. 

Unfortunately, emerging technology could render some of these mitigations ineffective. As recent work has observed~\cite{marschalek2018wolf}, ransomware authors share many of the challenges faced by \emph{confidential cloud computing}~\cite{sabt2015trusted}, which seeks to protect applications deployed on remote cloud infrastructure from a malicious cloud provider. To solve these challenges, confidential cloud computing leverages emerging support on commodity hardware for trusted execution environments~(TEEs), such as Intel SGX~\cite{costan2016intel,mofrad2018comparison}. SGX TEEs, commonly known as \emph{enclaves}, allow parts of an application to be executed securely irrespective of the rest of the system \cite{sabt2015trusted}. As has been shown, it is also possible for malware to hide malicious activities within enclaves to evade anti-virus systems~\cite{marschalek2018wolf}. However, previous studies have not considered the impact of enclaves on the complete lifecycle of ransomware cryptographic key management, and the practicality and performance overhead of such a ransomware variant is unclear. Given the damage ransomware causes, it is important to understand what new advantages such ransomware might provide to attackers, whether anti-malware mechanisms in existing enclave technology are robust against them, and whether additional anti-ransomware mitigations might be required. 

We perform an in-depth study of the potential implications of enclave-enabled ransomware using \RansomwareName{}, a proof-of-concept family of ransomware variants that leverages SGX to manage ransomware cryptographic keys on a victim's machine. \RansomwareName{}, once executed, creates an enclave that it uses to securely generate asymmetric key pairs. \RansomwareName{} releases the private key only when a ransom payment transaction from the Bitcoin blockchain is verified by the enclave. 

In contrast to previous work, we analyse the security and practicality of \RansomwareName{} with respect to all phases of the ransomware lifecycle. During the key generation and encryption phases, we show that there is a design-space whereby to avoid triggering network monitoring software a ransomware author may skip SGX remote attestation and avoid command and control (C2) server communication before encryption completes, but that this prevents dynamic loading of encrypted ransomware code to the enclave. Similarly, during the key release phase, we use \RansomwareName{} to show that enclaves open up a design-space of new key release schemes, each with different trade-offs for the attacker between security, reliability, and operational overhead. Of particular concern are fully autonomous or \emph{trustless} key release schemes, which do not require the victim to trust the attacker to release the decryption key after a ransom is paid. This guarantee could increase a victim's willingness to pay a ransom and by extension attacker profitability. 

\textbf{Contributions.} In summary, this paper makes the following contributions:
\begin{enumerate}[noitemsep,nolistsep]

    \item{We analyse key management schemes of existing ransomware variants and, based on our findings, derive requirements for a successful attack~(\S\ref{section3}).}
    \item{We introduce \RansomwareName{}, a proof-of-concept SGX-based ransomware that generates keys in an enclave to protect them from disclosure~(\S\ref{designsection}), and analyse its performance.}
    \item{We propose three blockchain-based \RansomwareName{} key release schemes and show how avoiding interaction with the attacker post-infection could potentially make the ransomware operation more profitable for attackers~(\S\ref{keyreleasesection}).}
    \item{We analyse mitigations to defend against SGX-based ransomware attacks~(\S\ref{discussionsection}).}
\end{enumerate}

\section{Background}
\thispagestyle{empty}
This section provides background on ransomware encryption mechanisms, and
a summary of relevant features of Intel SGX.

\mypar{Ransomware Encryption Schemes} Modern ransomware are often classified into three types based on the type of encryption employed. \emph{Symmetric encryption ransomware} uses one key for both encryption and decryption and allows faster file encryption, resulting in shorter time to complete the attack. This also reduces the chances of the attack being discovered by requiring less processing power for computations. \emph{Asymmetric encryption ransomware} uses asymmetric cryptography, with the public key used to encrypt files and the associated private master key to decrypt them. The encryption process is a resource-intensive task and slower in comparison to symmetric key encryption due to the overhead of public-key encryption.
However, it is more secure since the encryption process can be completed using only the public key thus allowing the private key to be kept secret. Finally, \emph{hybrid encryption ransomware} incorporates both symmetric encryption and asymmetric encryption. It uses symmetric encryption to encrypt the user files quickly. The symmetric key is then encrypted using asymmetric encryption. Hybrid encryption is often used by newer strains of ransomware, and reaps the benefits of both symmetric encryption (speed) and asymmetric encryption (security) \cite{bhardwaj2017ransomware,kok2019ransomware}.

\mypar{Intel SGX} Intel SGX is a security enhancement to recent Intel CPUs that provides a trusted execution environment (TEE). SGX TEEs, known as \emph{enclaves}, protect the integrity and confidentiality of applications handling sensitive data such as cryptographic keys. The SGX architecture enables an application to instantiate one or more enclaves such that enclave code and data is protected from the OS and hypervisor~\cite{costan2016intel,sinha2015moat}, and even from an attacker with physical access to the victim's computer \cite{mckeen2013innovative}.

Enclave code must execute in user space, and is not able to execute system calls or access secure peripherals. As such, applications are divided into two parts, a protected enclave (trusted code) and an unprotected part (untrusted code) which handles communication between the OS and enclave. The processor transparently encrypts and integrity protects enclave data whenever it leaves the CPU. The Intel SGX SDK provides a function call-like abstraction for entering and exiting an enclave. Calls into the enclave are referred to as \texttt{ECALL}s (enclave entry call), and calls from the enclave to outside as \texttt{OCALL}s (outside call)~\cite{mckeen2013innovative,costan2017secure,shepherd2017establishing}.

The SGX architecture includes a sealing capability, which allows data to be stored on persistent storage in an encrypted form. The sealing key is generated using a Key Derivation Function (KDF) \cite{costan2017secure} from the embedded base data seal key, which is fused into each processor during manufacturing by Intel. The sealing key is provided by the processor to the enclave, and only the enclave that sealed the data can later unseal it. SGX also supports remote attestation to allow a remote party to verify an enclave and establish a secure channel to it. To launch a production enclave in SGX version 1 (SGXv1), SGX required either the enclave binary or its author to be registered with Intel. In SGXv2, Intel added a flexible launch control capability, allowing the platform owner to bypass Intel as an intermediary in the enclave launch process \cite{schwarz2019practical,intel64and}.

\section{Ransomware Key Management}
\label{section3}
In this section, we discuss how ransomware key management schemes have evolved over the years. We then analyse the weaknesses of existing schemes in the face of modern anti-ransomware techniques, and propose five requirements of an attacker for a ransomware design. Notation used in our analysis and the remainder of the paper is as follows: (i) $\mathit{AK_{prv}}$ (attacker's private key); (ii) $\mathit{AK_{pub}}$ (attacker's public key); (iii) $\mathit{VK_{prv}}$ (victim specific private key); (iv) $\mathit{VK_{pub}}$ (victim specific public key); (v) $\mathit{EK_{v(f)}}$/$\mathit{EK_{v}}$ (symmetric key for victim $\mathit{v}$'s file $\mathit{f}$ --- index $\mathit{f}$ omitted where not relevant); (vi) $\mathit{nonce}$ (pseudo-random number).

\subsection{The Evolution of Ransomware Key Management}
\label{evolution}
At the heart of any ransomware operation lies key management: if poorly implemented, it can potentially leak encryption keys to the victim and threaten the entire campaign \cite{conti2018economic}. Table \ref{table1} summarises our analysis of key management schemes in some of the most prevalent ransomware families. Our analysis extends several previous analyses \cite{wyke2015current,berrueta2019survey,halderman2009lest}, in particular, we also show existing core ransomware functionalities and match them against a set of requirements (formally defined in \S\ref{requirements}) that must be met by ransomware variants.

\begin{table}[t]
\caption{Ransomware key management requirements}
\label{table1}
\centering
\renewcommand{\arraystretch}{1.1}
\scalebox{0.85}{
\begin{tabular}{|p{15mm}|p{18mm}|p{10mm}|p{5mm}|p{5mm}|p{5mm}|p{5mm}|p{6.8mm}|}
\hline
\multirow{2}{*}{\textbf{\begin{tabular}[c]{@{}l@{}}Encryption\\ Method\end{tabular}}} & \multirow{2}{*}{\textbf{Families}} & \multirow{2}{*}{\textbf{Release}} & \multicolumn{5}{c|}{\textbf{Ransomware RQ (\S\ref{requirements})}} \\ \cline{4-8}

& & & \centering \textbf{RQ1} & \centering \textbf{RQ2} & \centering \textbf{RQ3} & \centering \textbf{RQ4} & \textbf{ RQ5} \\ \cline{1-8}
& Apocalypse        &   \centering{2016}    & \centering{\ding{55}} & \centering{\ding{55}} & \centering{\ding{55}} & \centering{\ding{51}} & \centering\ding{55} \tabularnewline 
& Jigsaw            &   \centering{2016}    & \centering{\ding{55}} & \centering{\ding{55}} & \centering{\ding{55}} & \centering{\ding{51}} & \centering\ding{55} \tabularnewline 
                        \centering \multirow{-3}{*}{Symmetric}
& Razy              &   \centering{2016}    & \centering{\ding{55}} & \centering{\ding{55}} & \centering{\ding{55}} & \centering{\ding{51}} & \centering\ding{55} \tabularnewline \hline 

& CryptoLocker      &   \centering{2014}   & \centering{\ding{51}} & \centering{\ding{51}} & \centering{\ding{55}} & \centering{\ding{55}} & \centering\ding{55} \tabularnewline
& CBT-Locker        &   \centering{2014}   & \centering{\ding{55}} & \centering{\ding{55}} & \centering{\ding{55}} & \centering{\ding{51}} & \centering\ding{55} \tabularnewline
& CryptoDefense    &   \centering{2015}   & \centering{\ding{55}} & \centering{\ding{51}} & \centering{\ding{55}} & \centering{\ding{55}} & \centering\ding{55} \tabularnewline
                        \centering \multirow{-3}{*}{Asymmetric}
& PetrWrap          &   \centering{2016}    & \centering{\ding{51}} & \centering{\ding{55}} & \centering{\ding{55}} & \centering{\ding{51}} & \centering\ding{55} \tabularnewline
& Unlock92          &   \centering{2016}    & \centering{\ding{55}} & \centering{\ding{55}} & \centering{\ding{55}} & \centering{\ding{51}} & \centering\ding{55} \tabularnewline \hline
 
& CryptoWall        &   \centering{2014}    & \centering{\ding{51}} & \centering{\ding{51}} & \centering{\ding{55}} & \centering{\ding{55}} & \centering\ding{55} \tabularnewline 
& TorrentLocker     &   \centering{2014}    & \centering{\ding{55}} & \centering{\ding{55}} & \centering{\ding{55}} & \centering{\ding{51}} & \centering\ding{55} \tabularnewline 
& TeslaCrypt        &   \centering{2015}    & \centering{\ding{55}} & \centering{\ding{55}} & \centering{\ding{55}} & \centering{\ding{51}} & \centering\ding{55} \tabularnewline 
& Cerber            &   \centering{2016}    & \centering{\ding{55}} & \centering{\ding{51}} & \centering{\ding{55}} & \centering{\ding{51}} & \centering\ding{55} \tabularnewline
& Locky             &   \centering{2016}    & \centering{\ding{51}} & \centering{\ding{51}} & \centering{\ding{55}} & \centering{\ding{55}} & \centering\ding{55} \tabularnewline
                        \centering \multirow{-3}{*}{Hybrid}
& Petya             &   \centering{2016}    & \centering{\ding{51}} & \centering{\ding{51}} & \centering{\ding{55}} & \centering{\ding{55}} & \centering\ding{55} \tabularnewline
& RYUK              &   \centering{2017}    & \centering{\ding{51}} & \centering{\ding{55}} & \centering{\ding{55}} & \centering{\ding{55}} & \centering\ding{55} \tabularnewline
& WannaCry          &   \centering{2018}    & \centering{\ding{55}} & \centering{\ding{55}} & \centering{\ding{55}} & \centering{\ding{51}} & \centering\ding{55} \tabularnewline
& EKING             &   \centering{2020}    & \centering{\ding{51}} & \centering{\ding{55}} & \centering{\ding{55}} & \centering{\ding{51}} & \centering\ding{55} \tabularnewline \hline   
                        \centering{TEE}
& \RansomwareName{} &   \centering{2021} & \centering{\ding{51}} & \centering{\ding{51}} & \centering{\ding{51}*} & \centering{\ding{51}} & \centering\ding{51} \tabularnewline
\hline

\end{tabular}}
\begin{flushright}\footnotesize{* Only in proactive variant}\end{flushright}
\end{table}

{
\begin{figure}[b]
    \centering
        \includegraphics[scale=.830]{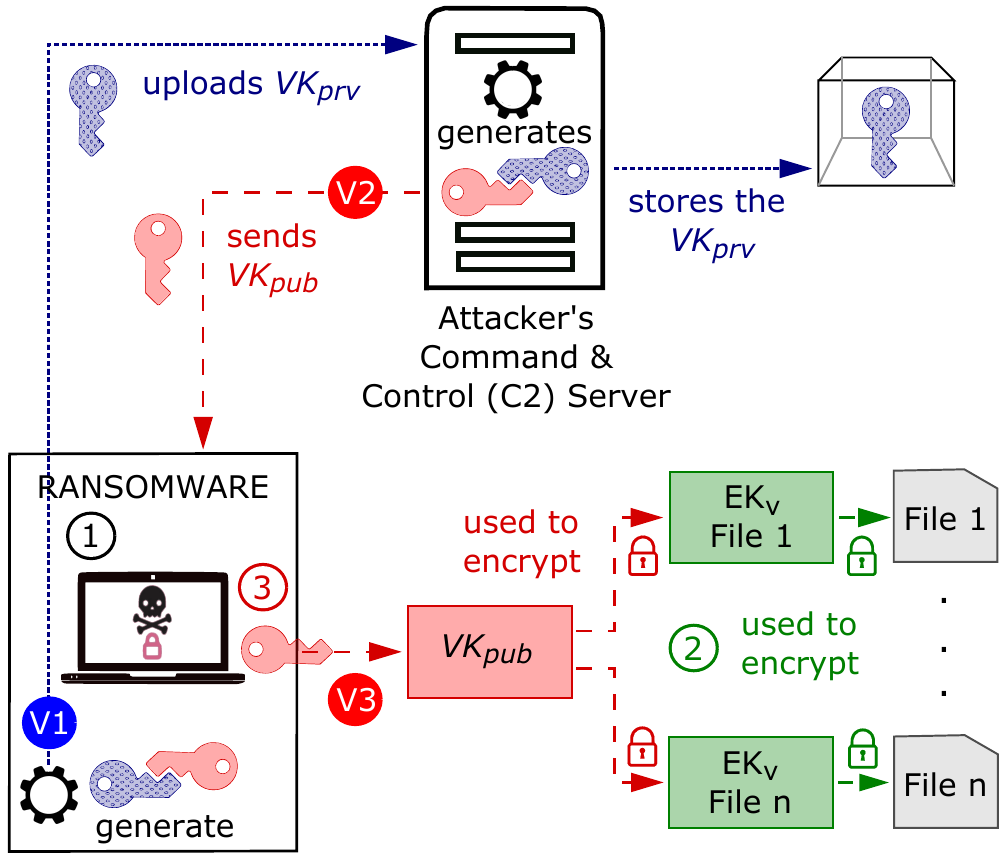}
    \caption{Current key management process of a typical ransomware}
    \label{fig:1}
\end{figure}
}

Our research shows that the majority (60\%) of ransomware variants rely on an active Internet connection on the victim's machine to exchange keys with a command-and-control~(C2) server~\cite{bajpai2018key,berrueta2019survey}. Additionally, 40\% of ransomware variants (e.g., WannaCry) generate keys on a victim's untrusted machine, thus leaving them exposed and vulnerable to recovery \cite{berrueta2019survey,halderman2009lest}. Older strains of ransomware, such as Apocalypse and CryptoLocker, use either symmetric key or asymmetric key encryption. Newer strains, such as CryptoWall, TeslaCrypt, WannaCry and RYUK often opt for a hybrid encryption method. As Figure \ref{fig:1} shows, there are three variants, {\large \circled[text=white,fill=blue, line width=-1mm]{\small V1}}, {\large \circled[text=white,fill=red, line width=-1mm]{\small V2}}, and {\large \circled[text=white,fill=red, line width=-1mm]{\small V3}} for key generation. In {\large \circled[text=white,fill=blue, line width=-1mm]{\small V1}}, ransomware, once executed ({\large \textcircled {\small 1}}), generates $\mathit{VK_{prv}}$ and $\mathit{VK_{pub}}$ locally on the victim's machine while sending the $\mathit{VK_{prv}}$ to the attacker's C2 server. In {\large \circled[text=white,fill=red, line width=-1mm]{\small V2}}, the ransomware generates the key pair on the C2 server, sending the $\mathit{VK_{pub}}$ to the victim's machine while safely storing the $\mathit{VK_{prv}}$. A unique $\mathit{EK_{v}}$ is generated for each victim file to be encrypted ({\large \textcircled{\small 2}}), and all $\mathit{EK_{v}}$ are then encrypted with the $\mathit{VK_{pub}}$ ({\large \textcircled{\small 3}}).

\subsection{Ransomware Key Management Requirements}
\thispagestyle{empty}
\label{requirements}
We define a set of desirable requirements from an attacker’s perspective that
modern ransomware must fulfil to achieve its primary goal successfully, i.e.
encrypt victim's data, secure the $\mathit{VK_{prv}}$ and hold the data hostage
until the ransom is paid. As we will see however, some of these requirements are
potentially in conflict with each other. 

\mypar{RQ1: Secure asymmetric key generation} Ransomware must ensure that the
$\mathit{VK_{prv}}$ needed to decrypt individual $\mathit{EK{v}}$ is not exposed
to the victim at any point before the ransom is paid. Many attackers prefer
setting up their own trusted C2 server to generate, distribute and securely
store $\mathit{VK_{prv}}$ as it provides some measure of protection against the
arbitrary key recovery. However, this setup requires an outbound connection before beginning the encryption process \cite{dargahi2019cyber,bajpai2018key}, which could be blocked/inspected by network monitoring tools \cite{morato2018ransomware,cusack2018machine}. In addition, if the server's location is compromised, it could potentially be taken down by law enforcement \cite{TheNoMor7:online}, thus exposing $\mathit{VK_{prv}}$ \cite{mims2017botnet}.
\label{R1}

\mypar{RQ2: Unique key per victim} Ransomware must generate a unique $\mathit{VK_{prv}}$ for each victim to maximise the ransom payment received from victims. Some attackers deploy ransomware {\large \circled[text=white,fill=red, line width=-1mm]{\small V3}}, such as IEncrypt, with an embedded $\mathit{VK_{pub}}$ to eliminate outbound connections. However, using one global $\mathit{VK_{prv}}$ could lead to victims collaborating and sharing the master key \cite{eviloffspring2016}. 
\label{R2}

\mypar{RQ3: Secure symmetric key generation} Ransomware in a hybrid encryption setting must ensure that the unique $\mathit{EK_{v}}$ used to encrypt each victim file are not exposed to the victim at any point.
\label{R3}

\mypar{RQ4: No post-exploitation connections}
To evade network monitoring software, Ransomware must avoid outbound connections to a C2 to retrieve the private keys or payload, or to an attestation service, until the victim's files have been encrypted. \label{R4}

\mypar{RQ5: Trustworthy key release} Ransomware must employ a secure and
reliable key release scheme post-attack to incentivise victims to co-operate and
pay the requested ransom \cite{Cartwright2019ToPO}. The scheme should also avoid
unnecessary operational overhead for the attacker (e.g. to maintain an online
presence, or manage large numbers of decryption keys)\label{R5}. Some attackers
request victims to send the encrypted $\mathit{VK_{prv}}$ along with the ransom
payment to avoid exposing the $\mathit{AK_{prv}}$. However, keys are still
generated on the victim's machine leaving them vulnerable.

\section{Threat Model}
In our threat model we assume the malware distribution and installation stages
have been successfully performed. We assume the ransomware has the appropriate
privilege to start enclaves and issue syscalls. We assume that the victim's
machine supports Intel SGX and the SGX feature is enabled. We revisit the
plausibility of these assumptions in our discussion of mitigations in~\S\ref{discussionsection}. 

We consider two different types of target systems: (i) standard system equipped with an AV, however, lacking any specific anti-ransomware functionality; (ii) advanced system equipped with an up-to-date AV that also includes an anti-ransomware module and can also capture cryptographic keys generated locally (e.g., PayBreak). In all cases, we assume that a victim cannot break the hardware-enforced security of SGX to obtain enclave-specific keys (e.g., sealing keys). We also assume that Intel would not release the key embedded into each processor during the manufacturing of the SGX-capable CPU \cite{anati2013innovative}.

\section{\RansomwareName{} Key Generation and Encryption}
\thispagestyle{empty}
\label{designsection}
As a basis for security and performance analysis of enclave-enhanced ransomware, we present \RansomwareName{}, a family of ransomware that securely manage their cryptographic keys using an enclave. In this section we describe
\RansomwareName{}'s key generation and encryption phase, and how conflicts
between the requirements introduced in \S\ref{requirements} motivate two
different \RansomwareName{} variants. We defer detailed discussion of
trustworthy key release~(\hyperref[R5]{RQ5}) to \S\ref{keyreleasesection}.

{
\begin{figure}[H]
    \centering
        \includegraphics[scale=.799]{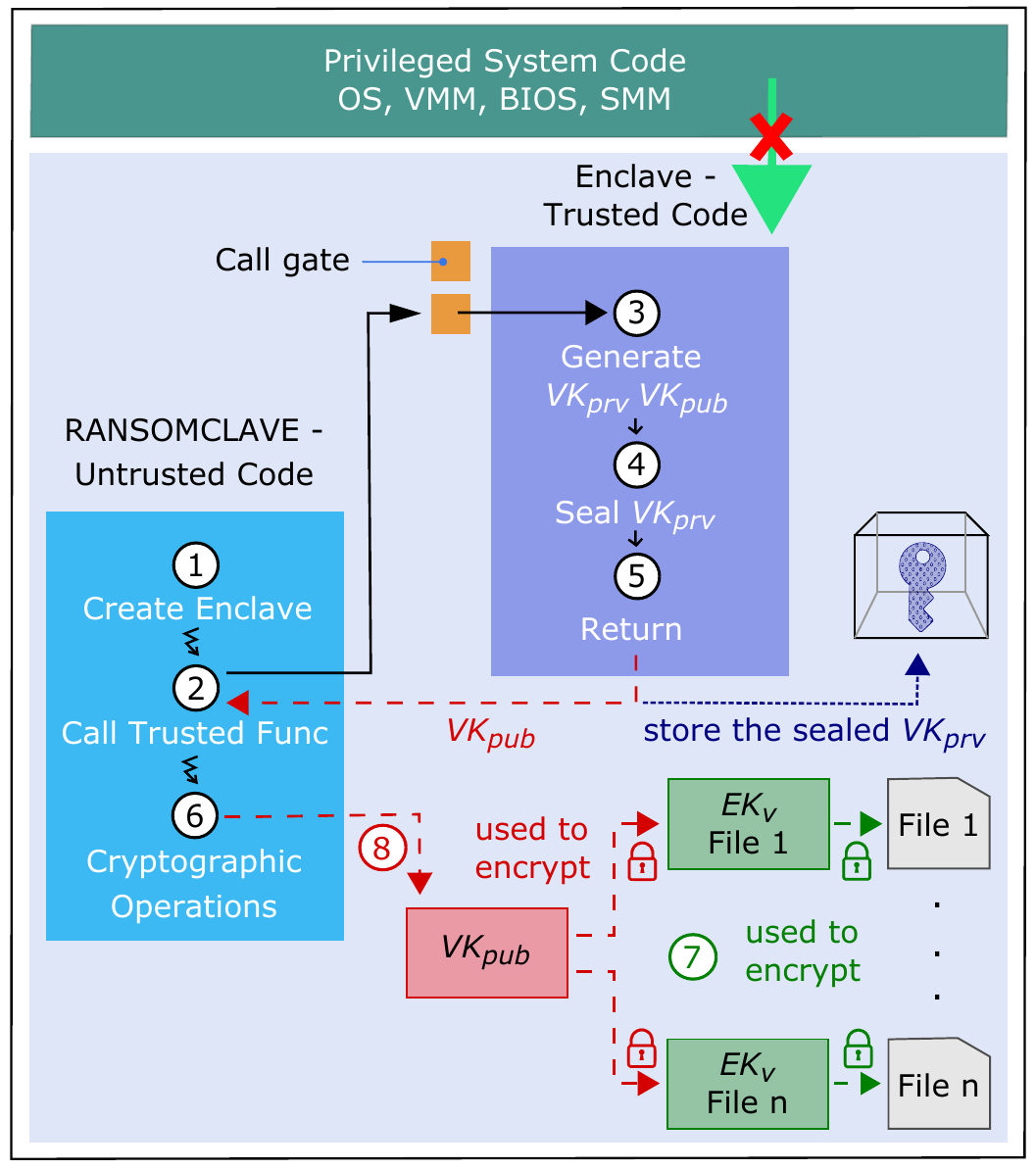}
    \caption{Key management and execution flow of \RansomwareName{} (steps
		{\large \textcircled{\small 7}}-{\large \textcircled{\small 8}} shown for reactive
		\RansomwareName{})}
    \label{fig:2}
\end{figure}
}

\RansomwareName{}'s key generation and encryption can be split into two stages. The first stage, asymmetric key generation (Figure~\ref{fig:2}, steps {\large \textcircled{\small 1}}-{\large \textcircled{\small 5}}), is common to both \RansomwareName{} variants. The second stage, symmetric key generation and encryption, differs between variants~(Figure~\ref{fig:2}, steps {\large \textcircled{\small 7}}-{\large \textcircled{\small 8}}).

\subsection{Asymmetric Key Generation}
\label{asymmetric}
As shown in Figure~\ref{fig:2}, the \RansomwareName{} malware consists of both
untrusted code (host application) and trusted code (enclave).
\RansomwareName{} asymmetric key generation happens primarily inside the enclave. Once
the malicious payload is executed, \RansomwareName{}'s untrusted code creates
the enclave {\large \textcircled{\small 1}}. The untrusted code then performs
an \texttt{ECALL} to enter the enclave and initiate key generation and
encryption {\large \textcircled{\small 2}}. On entering the enclave for the
first time, \RansomwareName{} generates a victim specific asymmetric key pair
$\mathit(VK_{prv}, VK_{pub})$ {\large \textcircled{\small 3}}.
\RansomwareName{} then creates an enclave-specific seal key using the
\texttt{EGETKEY} instruction, and seals $\mathit{VK_{prv}}$ using the seal key
{\large \textcircled{\small 4}}. Finally, the enclave performs an
\texttt{OCALL} to send the unique $\mathit{VK_{pub}}$ and sealed
$\mathit{VK_{prv}}$ to untrusted code {\large \textcircled{\small 5}}. Both
keys are then stored outside the enclave on the victim's system.

\subsection{Symmetric Key Generation and Encryption}
\label{sec:symmetric}
\RansomwareName{}'s symmetric key generation and encryption stage is similar to other modern ransomware that use a hybrid scheme for improved performance.
However, we introduce two different \RansomwareName{} variants for managing
symmetric keys, reactive and proactive, each focused on the capabilities of the target system.

\myparr{Reactive \RansomwareName{}:} This variant generates symmetric keys and
encrypts the victim's files \emph{outside} the enclave. It targets a
reactive victim who is unlikely to detect the ransomware and capture symmetric
keys exposed in untrusted memory (\hyperref[R3]{RQ3}) before encryption completes. However, encrypting files outside the enclave
avoids additional data movement overheads, allowing encryption to complete more quickly and thus serving as a baseline for performance comparison with other variants. 

Figure~\ref{fig:2} shows the untrusted code generating a unique
256-bit AES symmetric key ($\mathit{EK_{v(f)}}$) for each file $\mathit{f}$
before the encryption process starts~({\large \textcircled{\small 7}}). After
encryption of file $\mathit{f}$ completes, each $\mathit{EK_{v(f)}}$ is
encrypted using $\mathit{VK_{pub}}$ ({\large \textcircled{\small 8}}) and then
the plaintext $\mathit{EK_{v(f)}}$ is deleted. Only the encrypted version of
$\mathit{EK_{v(f)}}$ remains after the encryption process. Steps ({\large
\textcircled{\small 7}}) and ({\large \textcircled{\small 8}}) repeat until all
the victim files have been encrypted. The benefit of using per-file symmetric encryption keys is that the $\mathit{EK_{v(f)}}$ for files already encrypted will not be recoverable (since the original $\mathit{EK_{v}}$ is deleted after encrypting each file) even if the ransomware attack is discovered and stopped mid-execution. 

\myparr{Proactive \RansomwareName{}:} This variant assumes a proactive victim
who is vigilant and potentially able to start capturing cryptographic keys soon
after the ransomware is executed \cite{davies2020evaluation}~It generates $\mathit{VK_{prv}}$, $\mathit{VK_{pub}}$ and each
$\mathit{EK_{v(f)}}$ inside the enclave. In addition, the ransomware also
encrypts each victim file inside the enclave. By generating the $\mathit{EK_v}$
and encrypting files inside an enclave, the proactive variant mitigates some of
the security weaknesses of the reactive variant, i.e. exposure of the
$\mathit{EK_{v}}$ in memory (\hyperref[R3]{RQ3})

Although the proactive variant comes closer to meeting \hyperref[R3]{RQ3}, we
note that by default SGX does not provide confidentiality to enclave code. As a
result, a proactive victim with sufficiently advanced AV could inspect the
malware code and detect \RansomwareName{} given an appropriate signature.
Although an additional \RansomwareName{} variant with private dynamic code
loading is possible~\cite{dynsgx}, it would need to communicate with a C2 server
to download an encrypted malicious payload (violating~\hyperref[R4]{RQ4}). For
similar reasons, we do not consider using remote attestation to verify the
enclave.

Another potential drawback of the proactive variant is its performance, since it
must transfer victim files to the enclave from the host file system. Performing
file related system calls from the enclave requires additional costly enclave
transitions, potentially slowing the overall encryption process and jeopardising
the attacker's primary goal of holding the victim's data hostage quickly.
However, previous work has shown that asynchronous system calls can be used to
mask much of this overhead~\cite{arnautov2016scone}. We evaluate the overhead of
the proactive variant with this optimization in \S\ref{section7}.

\section{\RansomwareName{} Key Release}
\thispagestyle{empty}
\label{keyreleasesection}
Once \RansomwareName{} has successfully encrypted the victim's files and presented the victim with a ransom note, the next phase of its operation is key release. As discussed in \S\ref{evolution}, existing key release schemes introduce additional security risks and operational overhead for the attacker. Furthermore, they require the victim to trust the attacker to release the decryption key after a ransom is paid.

In this section we show how \RansomwareName{} enables more robust key release schemes in comparison to existing ransomware. We next explore three alternative \RansomwareName{} key release schemes, each providing different trade-offs between the security of the scheme, the operational overhead of the attacker for managing decryption keys, and the likelihood the victim will pay the ransom\footnote{In all these schemes, for simplicity, we assume that, once the private key has been released, the victim is able to decrypt the files autonomously, e.g. by using a decryptor that takes as input the private key and that is part of \RansomwareName{} or provided by the attacker independently.}.

\subsection{Blockchain with Online Attacker} Our first \RansomwareName{} key release scheme leverages a public blockchain (e.g. Bitcoin) and digital signatures to exchange metadata such that \RansomwareName{} will only unseal $\mathit{VK_{prv}}$ after the ransom has been paid.

\mypar{Security} From the attacker's perspective, the advantage of this approach is its security, since the attacker can independently check a ransom payment exists on the blockchain. The disadvantages are that it requires the attacker to maintain an online presence to monitor the blockchain, increasing operational overhead, and it requires the victim to actually trust the attacker to release the key after a payment has been made (as with current ransomware). We note that, unlike the next two schemes proposed later in this section, a variant of this scheme could potentially be used in combination with existing (non SGX-based) hybrid encryption ransomware.

\mypar{Overview} In this scheme, the victim is required to first make a ransom payment on the Blockchain that contains the following metadata in the transaction: $\mathit{AK_{pub}}$, $\mathit{VK_{pub}}$, and a $\mathit{nonce}$. Additionally, \RansomwareName{} derives a bitcoin wallet address based on the $\mathit{AK_{pub}}$ embedded in it and generates a random $\mathit{nonce}$ unique to the \RansomwareName{} enclave with high probability. The $\mathit{nonce}$ is used as input to a cryptographic hash function digitally signed later on the attacker's machine once the attacker notices the ransom payment. ($\mathit{AK_{prv}}$, $\mathit{AK_{pub}}$) key pair is generated/stored on the attacker's machine while ($\mathit{VK_{pub}}$, $\mathit{VK_{prv}}$) is generated inside an enclave and stored ($\mathit{VK_{prv}}$ as sealed) on the victim's machine. The victim next makes a ransom payment through Bitcoin \cite{blockchain_survey}. Once the attacker notices a new payment is on the blockchain, he creates a new Bitcoin transaction providing some uniquely identifiable piece of metadata, such as the $\mathit{nonce}$ signed using ($\mathit{AK_{prv}}$). \RansomwareName{} validates the authenticity and integrity of this data using the embedded $\mathit{AK_{pub}}$. If the verification is successful, the encrypted $\mathit{VK_{prv}}$ is unsealed and released to the victim. 

\mypar{Key Release Steps} The blockchain and digital signature payment verification scheme is illustrated in Figure \ref{fig:5}. We next describe the steps involved in detail:

{
\begin{figure*}[h!]
    
    \centering
        \includegraphics[scale=.800]{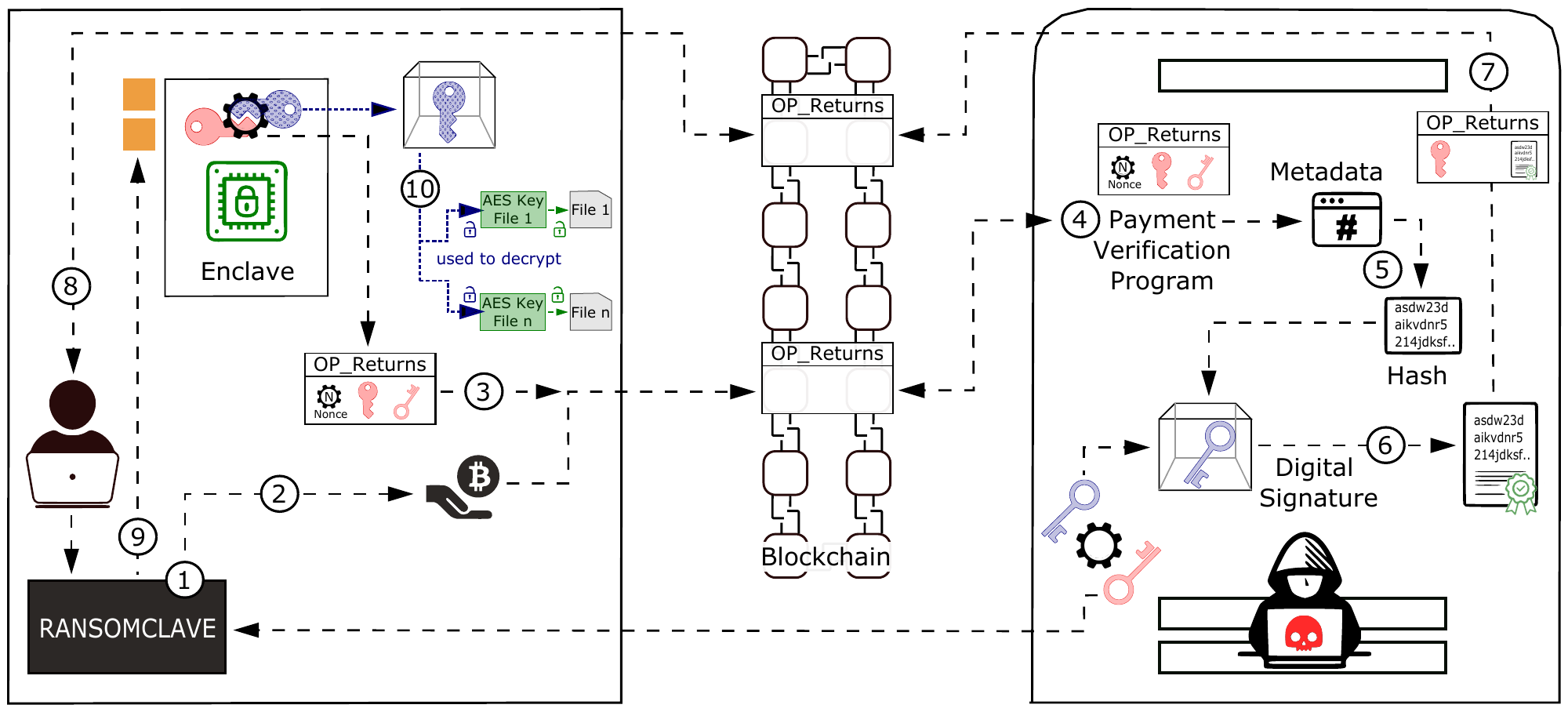}
    \caption {Key Release Scheme \#1: Blockchain with Online Attacker Payment Verification Diagram}
    \label{fig:5}
\end{figure*}
}

\begin{myitemize}

\item The victim follows the payment instructions ({\large \textcircled{\small 1}}) found on the ransom note (which also contains the bitcoin wallet address and the $\mathit{nonce}$) and deposits the requested amount of bitcoin into the newly generated wallet address ({\large \textcircled{\small 2}}). The victim also includes additional metadata consisting of the $\mathit{AK_{pub}}$, $\mathit{VK_{pub}}$, and $\mathit{nonce}$ into the blockchain transaction using an \texttt{OP\_RETURN} opcode ({\large \textcircled{\small 3}})\footnote{We assume the victim is provided with detailed instructions to do so or uses an automated tool to perform the necessary payment steps autonomously and easily.}. The $\mathit{nonce}$ is generated and used along with $\mathit{VK_{pub}}$ to give the attack a unique identification number. This is essential as it ensures another victim cannot use a previously obtained signed message hash from the attacker to trick the enclave into unsealing the $\mathit{AK_{prv}}$. \texttt{OP\_RETURN} is an instruction in the Bitcoin scripting language that allows users to attach metadata to a transaction and save it on the blockchain. It is used to allow the victim and the attacker to exchange metadata required to instruct the enclave to unseal the $\mathit{VK_{prv}}$.

\item The attacker periodically monitors the blockchain ledger for the $\mathit{AK_{pub}}$ in transaction metadata using a payment verification program. This program searches for the \texttt{OP\_RETURNS} opcode in an output's \texttt{scriptPubKey}\footnote{A locking mechanism placed on an output to prevent others from spending the bitcoin.} ({\large \textcircled{\small 4}}).

\item When a transaction containing $\mathit{AK_{pub}}$ is found and the payment is verified, the attacker employs a cryptography library to create a message hash using the SHA-256 hash of the metadata provided by the victim ({\large \textcircled{\small 5}}). The hash is then encrypted with $\mathit{AK_{prv}}$ ({\large \textcircled{\small 6}}) using the standard RSA signature algorithm to obtain the signature (the RSA encrypted message hash). This process cryptographically binds the signed hash with the $\mathit{AK_{prv}}$, thus allowing \RansomwareName{} to verify the signature using the embedded $\mathit{AK_{pub}}$. A new transaction is uploaded on to the blockchain containing the signed hash and $\mathit{VK_{pub}}$ ({\large \textcircled{\small 7}}).

\item The victim periodically searches for new transactions containing the $\mathit{VK_{pub}}$ (e.g. on blockchain explorer websites, such as blockchain.com) as per instructions provided in the ransom note. Once a transaction is found, the victim downloads the signed hash and submits it to \RansomwareName{} along with the original metadata consisting of $\mathit{AK_{pub}}$, $\mathit{VK_{pub}}$ and $nonce$ ({\large \textcircled{\small 8}}).

\item \RansomwareName{} performs an \texttt{ECALL} ({\large \textcircled{\small 9}}) to enter the enclave which computes a hash of the original metadata. The algorithm then decrypts the message signature with the public key exponent of $\mathit{AK_{pub}}$ to obtain the hash computed by the attacker. The two hashes are compared and if they match, \RansomwareName{} considers the signature valid. \RansomwareName{} then unseals and releases $\mathit{VK_{prv}}$ to the victim. 

\end{myitemize}

\subsection{Enclave SPV Client}

\label{scheme2}
To eliminate the attacker's online presence, we next propose a fully \emph{autonomous} and \emph{offline attacker} (\emph{trustless}) solution that uses a Simplified Payment Verification (SPV) client inside the enclave to verify ransom payments on the blockchain. In this scheme, \RansomwareName{} is deployed with a lightweight blockchain client embedded in it. This allows \RansomwareName{} to successfully verify the victim's payment on the blockchain and release $\mathit{VK_{prv}}$ without inputs from the attacker.

\mypar{Security} The main advantage of this scheme is that it is fully autonomous. This reduces operational overhead for the attacker, who no longer needs to maintain an online presence. More importantly, full autonomy means the victim only needs to trust\footnote{We define trust as a utility function where we express implicit trust by paying would be an expectation that the victim receives something in return, i.e. the decryption key.} the enclave code, whose behaviour post-payment can be verified up front using attestation. This is a very attractive property for the attacker potentially, since it can increase the chances of infected victims paying the ransom. On the downside, this scheme has arguably weaker security. In fact, enclave interactions with a blockchain are potentially vulnerable to man-in-the-middle (MitM) attacks~\cite{loe2019you}. Since the victim has full control over the enclave's network connectivity, it can impersonate the blockchain to trick the enclave into accepting fake blocks. However, mining plausible fake blocks still incurs an economic cost for the victim\footnote{The estimated cost of mining a Bitcoin block in 2018 was between \$2k-\$20k depending on the cost of electricity, e.g. see: \href{ https://www.marketwatch.com/story/heres-how-much-it-costs-to-mine-a-single-bitcoin-in-your-country-2018-03-06}{https://www.marketwatch.com/story/heres-how-much-it-costs-to-mine-a-single-bitcoin-in-your-country-2018-03-06}.}. As we will show, in many cases it is practical for the attacker to increase this cost such that it is cheaper for the victim to pay the ransom.

\mypar{Overview} Currently, there are two primary methods of validating the blockchain as a client: full nodes and SPV clients. A full node stores a copy of the entire blockchain with all the transactions, from the genesis block to the most recently discovered block. In contrast, SPV clients can check if particular transactions are included in a block through block headers only (80 bytes per block) and a Merkle tree rather than downloading the entire block \cite{bunz2019flyclient,antonopoulos2014mastering}.

Our second \RansomwareName{} key release scheme embeds an SPV Client inside the enclave, together with an embedded hash of the most recent stable block at the time the attacker creates the malware. The SPV client inside the enclave then requests full nodes to supply it with the headers of all blocks since the block corresponding to the embedded hash, in addition to a Merkle path for the transaction containing the ransom payment. Once the SPV client is satisfied the payment has been made, the enclave releases $\mathit{VK_{prv}}$ to the victim. Although full nodes usually give stronger security than SPV clients, for \RansomwareName{} this is not the case. Since the victim has full control over the attacker's network connectivity, it can mine a new block on top of the current longest Bitcoin chain with a fake transaction containing a supposed ransom payment. The victim must simply execute the standard Bitcoin proof-of-work mining algorithm to find a suitable nonce value. Even if the malware contains a full node, the hash of the genesis block, and the most recent block at the time the malware is created, the enclave cannot distinguish subsequent blocks mined by the victim from blocks on the real Bitcoin blockchain. \RansomwareName{} therefore avoids the additional storage and synchronisation overhead of a full node using an SPV client.

\thispagestyle{empty}

Depending on the size of the ransom, it may be cheaper for the victim to mine a fake block than pay the ransom. As a counter-measure, the attacker can require the victim to provide a valid chain with $\mathit{n}$ additional blocks mined on top of the block containing the ransom payment transaction. The victim would then need to mine a fake \emph{chain} of length $\mathit{n}$, linearly increasing the cost to fake a ransom payment. However, this mitigation would also require ``honest'' victims to wait for $\mathit{n}$ blocks after their payment has been made, and thus may not be practical for very large ransoms if timely key release is important. In this strategy, the attacker's bitcoin address is embedded in the \RansomwareName{}  binary. Additionally, the victim must include the unique nonce generated by the enclave into the bitcoin transaction. \RansomwareName{} will use this metadata to verify the uniqueness of the transaction prior to unsealing the $\mathit{AK_{prv}}$.

\mypar{Key Release Steps} As shown in Figure \ref{fig:6}, \RansomwareName{} executes the SPV client inside an enclave which in turn connects to a full Bitcoin node. We next describe the steps in detail: (i) after \RansomwareName{} receives the payment transaction from the victim ({\large \textcircled{\small 1}}), it loads the sealed victim's wallet address into enclave and establishes a TCP connection ({\large \textcircled{\small 2}}) to the full Bitcoin node. (ii) the SPV client sends a $\mathit{getblock}$ message ({\large \textcircled{\small 3}}) to request a list of blockchain headers from the current block number and hash embedded in \RansomwareName{} when the attacker created it. (iii) the full node checks transactions and sends a $\mathit{merkleblock}$ message ({\large \textcircled{\small 4}}), which comprises of block headers of the blocks and the Merkle path of the matched transaction. (iv) the SPV client ({\large \textcircled{\small 5}}) uses the $\mathit{merkleblock}$ to verify that the ransom payment transaction is included in the blockchain and has more than six confirmations. If the verification is successful, then the $\mathit{VK_{prv}}$ is unsealed and released.

{
\begin{figure}[h!]
    \centering
        \includegraphics[scale=.800]{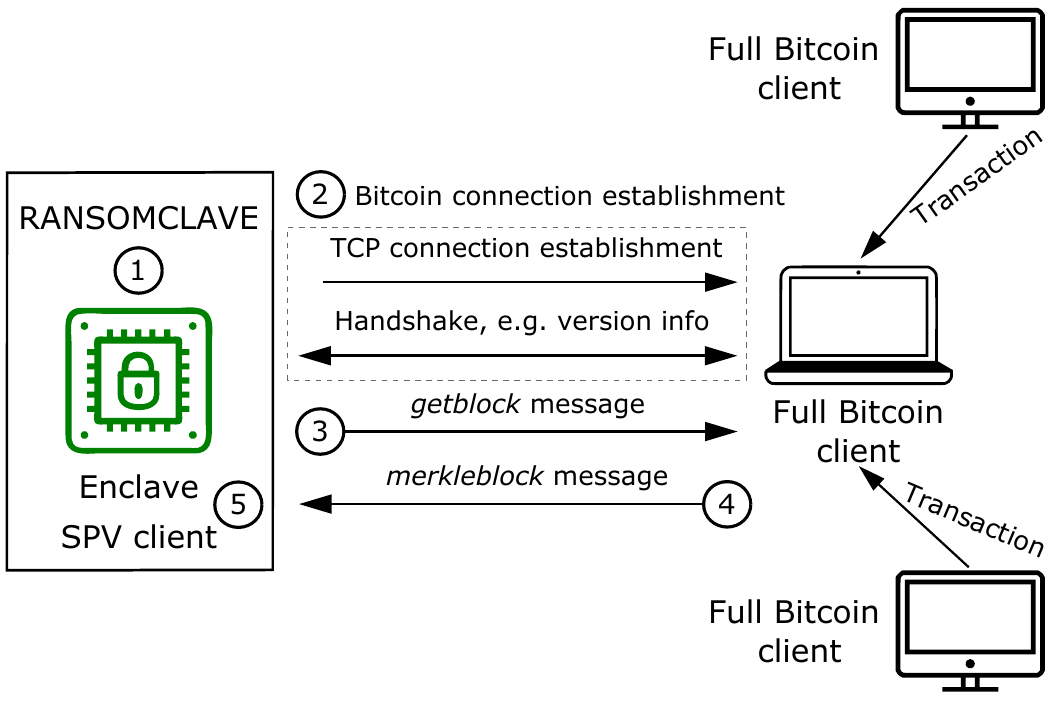}
    \caption{Key Release Scheme \#2: Enclave SPV Payment Client Verification Diagram (a revised version from \cite{gervais2014privacy})}
    \label{fig:6}
\end{figure}
}

\subsection{Blockchain Explorer Client}
To further improve the security of the SPV client scheme, we propose an alternative autonomous key release scheme that uses a third-party \emph{blockchain explorer service}\footnote{For example, Blockchain.com provides an explorer API at \href{https://www.blockchain.com/api}{https://www.blockchain.com/api}.}. Blockchain explorer services allow clients to retrieve transaction(s) relating to a particular bitcoin address over a secure HTTPS channel. 

\mypar{Security} This approach provides security against straightforward MitM attacks. To subvert it, the victim must convince the blockchain explorer service to provide fake data to the enclave (or pressure law enforcement to force them). Even if this is possible, the attacker can combine this scheme with the previous SPV client scheme such that the victim must still pay the cost required to mine plausible fake blocks. One downside of this approach is that it introduces an additional dependency on a third party service. If this becomes unavailable for any reason post-infection, even an ``honest'' victim will no longer be able to retrieve the decryption key $\mathit{VK_{prv}}$ from the enclave.

\mypar{Key Release Steps}
In this scheme, when \RansomwareName{} is executed, the enclave initiates a TLS connection with the Blockchain explorer API server. We describe the steps in detail: (i) Once the TLS Handshake is complete, the server's digital certificate is retrieved along with information about the Certificate Authority (CA) that issued it. The digital certificate contains the public key and the identity of the owner. Additionally, it is signed by the CA using its private master key. This allows the enclave to use a pre-seeded store of SSL certificates authorities' public keys embedded in the binary to verify integrity and authenticity of the retrieved certificate. (ii) Once a secure channel has been established, the enclave makes an API call to the Blockchain explorer API server for transactions and block data relating to the attacker's Bitcoin addresses. (iii) The server sends relevant data to the enclave. If a valid transaction with a minimum of six confirmations is found, then the enclave releases the $\mathit{VK_{prv}}$. Six confirmations is generally accepted for most transactions as it represents enough security to ensure the validity of the transaction.

\section{Implementation and Evaluation}
\thispagestyle{empty}
\label{section7}

We present a proof-of-concept implementation of \RansomwareName{} design. Our
prototype implementation operates similarly to WannaCry but executes inside an
Intel SGX enclave. The reactive variant of \RansomwareName{} is composed of
2,600 SLOC, while for the proactive variant, the SGX-LKL library OS is also needed as well as code restructuring for symmetric key generation.

\begin{algorithm}
\SetAlgoNoLine
\KwResult{Secret launchToken for the enclave, createEnclave and encryptFile}
Input: $\mathit{enclaveId}$\;
$\mathit{launchToken} \gets \mathit{secretKey()}$\;

\If{createEnclave(launchToken, enclaveId)}
{

 $\mathit{VK_{prv}}, \mathit{VK_{pub}}\gets \mathit{enclaveGenKey()}$\;
 $\mathit{VK_{prv}} \gets \mathit{seal(enclaveId}, \mathit{VK_{prv}})$\;
$\mathit{VK_{pub}} \gets \mathit{save(enclaveId}, \mathit{VK_{pub}})$\;

}
\Else
{
    $\mathit{abort()}$;
}

$\mathit{VK_{pub}} \gets importKey()$\;

 \For{file $\mathit{f}$ in getImportantFiles()}{

   $\mathit{EK_{v(f)}} \gets \mathit{generateSymKey()}$ \;

   $\mathit{encrF} \gets \mathit{encrypt(f}, \mathit{EK_{v(f)}})$\;
   $\mathit{encrSymKey}\gets \mathit{encrypt(EK_{v(f)}}, \mathit{VK_{pub}})$\;
   $\mathit{fConcat} \gets \mathit{append(encrSymKey, encrF)}$\;
   $\mathit{delete(f)}$\;
   $\mathit{save(fConcat)}$\;

}
$\mathit{displayRansomNote()}$\;
\caption{Setup and Execution of \RansomwareName{}}

\label{algo1}
\end{algorithm}

\mypar{Experimental Setup.} All our experiments were performed on Intel\textsuperscript{\textregistered} Dual-Core\textsuperscript{\texttrademark} i5-7360U CPU with 4GB of RAM, and 3 CPU cores at 2.30GHz. Our system ran Ubuntu 18.04.5 LTS 64-bit within VirtualBox 6.0. In addition, we performed our tests using Intel SGX drivers \cite{GitHubin94:online}, the Intel SGX SDK \cite{GitHubin5:online}, and the open-source SGX-LKL library OS~\cite{lsdssgxl64:online}.

\mypar{\RansomwareName{}} Algorithm \ref{algo1} describes the \RansomwareName{} program in reactive variant. In detail, when the malware is executed, its first task is to initiate the enclave using a launch token and get the buffer sizes required by the untrusted code to store the $\mathit{VK_{pub}}$ and sealed $\mathit{VK_{prv}}$. Once the memory is allocated, the untrusted code makes an \texttt{ECALL} by calling the \texttt{enclaveGenKey} function to generate ($\mathit{VK_{prv}}$, $\mathit{VK_{pub}}$) pair using 256-bit Elliptic Curve Cryptography (ECC). The enclave seals $\mathit{VK_{prv}}$ using the data seal key and returns both keys to the untrusted code. When the keys are in place, \RansomwareName{} initiates the encryption process by calling the \texttt{getImportanFiles} function, which starts searching for the victim's files in specified targeted folders and encrypting each file $\mathit{f}$ using a new $\mathit{EK_{v(f)}}$. Each $\mathit{EK_{v(f)}}$ is then encrypted using $\mathit{VK_{pub}}$ before it is permanently deleted. The encryption process is repeated until there are no new files to encrypt. Finally, a ransom note is displayed to the victim containing instructions on how to make the payment and unseal $\mathit{VK_{prv}}$ to restore the files.

\begin{algorithm}

\SetAlgoNoLine
\KwResult{Launch enclave, initiate TLS connection, get transaction data, verify and unseal $\mathit{VK_{prv}}$}
Input: $enclaveId$, $hostName$, $portNum$, $walletAddr$, $txId$\;
$launchToken \gets secretKey()$\;

\If{loadEnclave(launchToken, enclaveId)}
{
    $\mathit{SSL\_Library\_init()}$\;

    $\mathit{ctx} \gets \mathit{initCTX()}$\;
    $\mathit{server} \gets \mathit{openConnection(hostName, portNum)}$\;
    $\mathit{ssl} \gets \mathit{sslNew(ctx)}$\;
    $sslSetFd(ssl, server)$\;
    
    \If{(sslConnect(ssl) == FAIL)}
     {
        $\mathit{abort()}$\;
     }
    \Else
    {
        $\mathit{txResult} \gets \mathit{getTx(hostName, walletAddr, txId)}$\;
        \If{($\mathit{verify(txId)}$ == TRUE)}
        {
            $\mathit{{VK_{prv}}} \gets \mathit{unseal(enclaveId, {VK_{prv}})}$\;
            $\mathit{release({VK_{prv}})}$\;
        }
        \Else
        {
            $\mathit{print}$("Verification failed")\;
        }
    }
}
\caption{Enclave Launch and Key Release of \RansomwareName{}}

\label{algo2}
\end{algorithm}

\mypar{Proactive \RansomwareName{}} We extend our proof-of-concept
\RansomwareName{} to support the proactive variant using the open-source SGX-LKL
library OS \cite{lsdssgxl64:online}, which is designed to run existing
unmodified Linux binaries inside enclaves. SGX-LKL executes the ransomware
sample to perform cryptographic operations on sample victim data (provided as
part of a Linux disk image) inside an enclave. The in-enclave SGX-LKL library OS
provides transparent system calls (e.g. file and network I/O), user-level
threading, and signal handling. Reactive \RansomwareName{} is categorised as native since its cryptographic operations are performed in user-space similar to traditional ransomware operation.

\thispagestyle{empty}

\mypar{Blockchain Explorer Client} Algorithm \ref{algo2} describes the blockchain explorer services with offline attacker scheme we have implemented within \RansomwareName{}. When \RansomwareName{} is executed, the victim is asked to provide a bitcoin address along with the transaction identifier (\texttt{txId}) of the ransom payment. An enclave is then loaded using a launch token followed by an \texttt{OCALL} to initiate a TLS handshake with the Blockchain API server. The \texttt{getTx} message is sent to the server to request a data block linked to \texttt{txId} and wallet address (\texttt{walletAddr}) along with number of confirmations. If a valid transaction with at least six confirmations is received, then the enclave executes the \texttt{unseal} function to unseal and release $\mathit{VK_{prv}}$.

\mypar{Operational and Performance Evaluation}
We have tested the proof-of-concept \RansomwareName{} operational capabilities on four different SGX-enabled VMs, each configured to store different types of files as to mimic user's and organizational scenarios. In each run, \RansomwareName{} encrypted 200 files whose size ranged from few KB up to 250MB. After the encrypting operations, the user was shown a ransom note requesting a symbolic sum, and we  performed the necessary steps to perform key release.

\begin{figure}[ht]
    \centering
        \includegraphics[scale=.689]{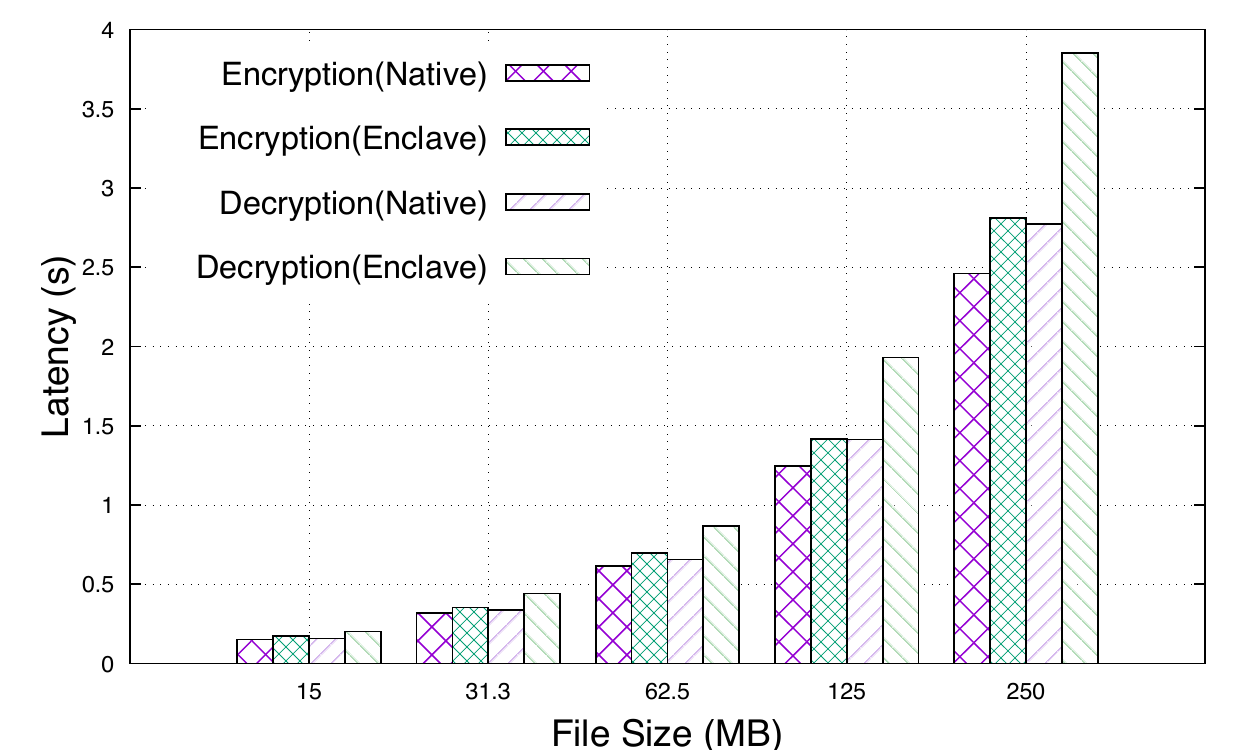}
    \caption{Cryptographic operation performance evaluation}
    \label{fig:10}
\end{figure}

Our performance evaluation compares the speed of cryptographic operations for the two \RansomwareName{} variants proposed in this paper, reactive (native) and proactive (enclave). Figure \ref{fig:10} shows the measurement of cryptographic operations on files whose size ranges from 15MB to 250MB: we can see that these operations in the enclave take longer to complete when compared to the native ransomware. An average of 12.76\% overhead for the encryption operation and 34.05\% in decryption operation is observed when the ransomware sample is executed inside an enclave. Additionally, the results also revealed that the decryption operation is slower for both the native and enclave ransomware. This could be potentially due to the implementation of AES in Cipher Block Chaining mode, which uses cipher in the inverse direction, thus resulting in slower decryption timings \cite{rogaway2011evaluation}. However, it is possible to achieve faster decryption on a system that supports parallel decryption, although this may result in using more memory due to loading multiple copies of the encrypted block.


\mypar{Key Capture} We tested \RansomwareName{}'s key management against popular memory forensics tools and techniques, such as Dumpit, Volatility, and PayBreak \cite{kolodenker2017paybreak}. Our experiments confirm it is possible to extract keys from memory for a reactive variant that generates symmetric keys outside of an enclave. However, it is not feasible for the proactive variant as all keys are generated inside an enclave.

\section{Discussion}
\thispagestyle{empty}
\label{discussionsection}
\label{section8}

\mypar{RansomClave Pre-Infection Mitigations}
Several methods can be applied to reduce \RansomwareName{}'s attack surface. In particular, given that the main focus of \RansomwareName{} is to provide robust key management, some existing mitigations against ransomware based on static analysis or in-memory analysis techniques~\cite{berrueta2019survey,moussaileb2018ransomware,kharaz2016unveil,morato2018ransomware,kolodenker2017paybreak,alam2020rapper,cusack2018machine,bajpai2020memory} can detect \RansomwareName{}. For instance, binary inspection of an unknown program could detect cyptographic calls \cite{grobert2011automated} or file
calls \cite{8424649}. In addition, as shown in \cite{bergeron2001static,schwarz2020malware}, it is also possible to detect unknown malicious code and functions in enclave binaries using existing static
analysis techniques, since SGX by default does not provide code confidentiality. Furthermore, a malicious program can only issue system calls through the host application: as such, software monitoring system calls could stop the malware before it can take any destructive actions. A user could also disable SGX functionality in the system BIOS when not in use, although this is undesirable. Lastly, system privilege is needed to create an enclave. As such a strict policy could be enforced on the end user machine to stop \RansomwareName{} execution.

\mypar{Launch Control} Intel SGX's launch control whitelist provides a mechanism to prevent wide-scale abuse of enclaves on systems where it is straightforward to gain admin privileges. However it requires Intel to proactively monitor and revoke whitelisted keys that have been compromised or abused, and for systems to regularly update their local copy of the whitelist. We note however that openness concerns around launch control have hindered upstream support for SGX in the Linux kernel. The recent availability of Flexible Launch Control on newer versions of SGX mitigates some of these openness concerns, but shifts the burden for securely managing the whitelist to the system owner, and potentially requires a secure-boot mechanism to protect the whitelist against an attacker with admin privileges. To the best of our knowledge, other enclave technologies (e.g. AMD SEV) do not have a hardware vendor-enforced launch control whitelist.

\mypar{RansomClave Post-Infection Mitigations} As shown by \cite{chen2019sgxpectre,nilsson2020survey}, SGX is vulnerable to software-based side-channel attacks that, according to Intel, can only be prevented by the developers themselves\footnote{E.g., see: \href{https://software.intel.com/content/www/us/en/develop/articles/intel-sgx-and-side-channels.html}{https://software.intel.com/content/www/us/en/develop/articles/intel-sgx-and-side-channels.html}}, i.e., Intel does not consider side channel attacks as part of the SGX threat model. Hence, victims infected with \RansomwareName{} could potentially extract ransomware's keys from inside of an enclave via side-channel attacks.

\mypar{Extensions}
We have considered \RansomwareName{} targeting multiple, distinct users. In an enterprise scenario, rather than creating enclaves on multiple hosts, \RansomwareName{} can set up a single internal SGX server to manage all the cryptographic keys of the infected hosts on a local network. In this instance, a clear advantage from an attacker's perspective is that it simplifies the key management, particularly if some hosts do not have SGX support. However, relying on a single machine to generate all keys could impact untrusted code's ability to promptly encrypt victim's files since the encryption key is not immediately available. Additionally, it also leads to a single point of failure for the attack. 

\mypar{Ethical Considerations}
This research has ethical concerns of dual use research as our findings could improve the key management of modern ransomware. However, by commenting on the theoretical risks of future SGX-based ransomware, we preemptively foster discussions on the possible risks of SGX-empowered malware that might have been underestimated \cite{schwarz2019practical}. We also omit implementation details to develop a fully-fledged SGX-based ransomware and key release scheme -- this would require cyber-criminals to perform a non-trivial integration task -- and outline mitigation improvements.

\section{Related Work}
\label{section9}
\textbf{SGX Malware}
 SGX-ROP is the first attack to enable an enclave malware to stealthily control its host application independent of the enclave API~\cite{schwarz2019practical}. SGX-ROP uses an Intel Transactional Synchronisation Extensions-based technique to construct a write-anything-anywhere primitive and a memory-disclosure primitive from inside an enclave. SGX-ROP can bypass ASLR, stack canaries, and address sanitiser, to run ROP gadgets in the host context, enabling practical enclave malware. However it does not evaluate the implications for key release. Marshalek
 discusses the possibility of ransomware running inside enclaves, but does not
 evaluate performance overhead or the implications for key
 release\cite{marschalek2018wolf}.
 
\mypar{Enclave Interaction with Blockchain} Authors \cite{goyal2017overcoming} proposes using blockchain to implement one-time programs via cryptographic obfuscation techniques. \cite{kaptchuk2019giving} expands on this idea by proposing a protocol that uses an enclave to facilitate secure state management for randomised multi-step computations. It also introduces the concept of combining the enclave with public ledgers to condition a program execution on the publication of particular messages on the ledger, and describes a practical set of applications leveraging this protocol. In contrast, our research provides an in-depth analysis of the trade-off and security advantage between different blockchain-enabled key release schemes.
 
\mypar{SGX-based Key Management Frameworks} Some research efforts
\cite{park2019sgx,cloud-micro-services} explore the application of SGX in a
cloud environment to minimise the disclosure of sensitive data to a third-party
hosting and providing the cloud services. Authors \cite{park2019sgx} proposes
key-management frameworks for data-centric networking where an SGX-based key
server, which supports remote attestation and key sealing, generates and manages
data encryption and decryption keys. Neither work addresses the implications of
SGX for ransomware key management.

\section{Conclusion}
\label{section10}

Based on an analysis of key management schemes of existing ransomware, we
observe that most ransomware strains either generate keys on the victim's
machine, thus leaving them vulnerable to memory key extraction techniques, or
require contact with a C2 server. However, recent work has raised the
spectre of future enclave-enhanced malware that can avoid such
mitigations using emerging support for hardware-enforced secure enclaves. Given
the damage ransomware causes, it is important to understand what new advantages
such ransomware might provide to attackers.

We implement a proof-of-concept called \RansomwareName{} to demonstrate the
practicality of such ransomware emerging in the near future. Our evaluation
shows that \RansomwareName{} can provide a practical key management solution
with minimal overhead to address known weaknesses in existing ransomware
strains. \RansomwareName{} also raises the possibility of autonomous
blockchain-based key release schemes that no longer require victims to trust the
attacker to release decryption keys or the attacker to maintain an online
presence, potentially increasing attacker profitability. We show however that a
variety of mitigation techniques are still possible for a security-aware victim.

\begin{acks}
\thispagestyle{empty}
This research of Alpesh Bhudia is supported by the EPSRC and the UK government as part of the Centre for Doctoral Training in Cyber Security at Royal Holloway, University of London (EP/P009301/1).
\end{acks}



\bibliographystyle{ACM-Reference-Format}
\bibliography{arxivpaper}


\end{document}